\documentclass{ws-procs9x6}
\def\deg{{$^{\circ}$}}
\def\Msun{\mbox{M$_{\odot}$}}
\def\bd17{\mbox{BD +17\deg 3248}}
\def\gtaprx{ \mathrel{  \vcenter{
                        \offinterlineskip \hbox{$>$}
                        \kern 0.3ex \hbox{$\sim$}    } } }
\def\ltaprx{ \mathrel{  \vcenter{
                        \offinterlineskip \hbox{$<$}
                        \kern 0.3ex \hbox{$\sim$}    } } }
\begin{document}

\title{
r-Process Abundance Signatures 
}

\author{John J. Cowan\footnote{\uppercase{A}lso at:
 \uppercase{D}epartment 
of \uppercase{A}stronomy  and \uppercase{M}cDonald \uppercase{O}bservatory, 
\uppercase{U}niversity of 
\uppercase{T}exas, \uppercase{A}ustin, \uppercase{T}\uppercase{X} 78712 
} }

\address{Department of Physics and Astronomy, \\
University of Oklahoma, \\ 
Norman, OK 73019,  USA\\ 
E-mail: cowan@nhn.ou.edu}

\author{Christopher Sneden}

\address{Department of Astronomy and McDonald Observatory, \\
University of Texas, \\ 
Austin, TX 78712, USA\\ 
E-mail: chris@verdi.as.utexas.edu}


\maketitle

\abstracts{
Abundance  observations	indicate the presence of
rapid-neutron capture (i.e., {\it r}-process) elements in old Galactic halo
and globular cluster stars.
These observations demonstrate that
the earliest generations of stars in the Galaxy, responsible
for neutron-capture synthesis and the progenitors of the halo stars,
were rapidly evolving.
Abundance comparisons among large numbers
of stars provide clues about the nature of neutron-capture  element synthesis both during the earliest times and throughout the history of the
Galaxy. In particular, these comparisons suggest differences in
the way the heavier (including Ba and above)
and lighter neutron capture elements are synthesized in nature. Understanding these differences will help to identify the astrophysical site (or sites) of
and
conditions in the {\it r}-process.
The abundance comparisons also demonstrate a
large star-to-star scatter in the neutron-capture/iron ratios at
low metallicities- which disappears with increasing [Fe/H]-
and suggests an early, chemically unmixed and inhomogeneous Galaxy.
The  very recent neutron-capture element observations
indicate  that the early phases of Galactic nucleosynthesis, and the associated
chemical evolution,  are  quite complex,
with the yields from different (progenitor)
mass-range stars contributing to  different chemical mixes.
Stellar abundance comparisons  suggest  a change from the {\it r}-process 
to the slow
neutron capture ({\it i.e.}, {\it s}-) 
process at  higher metallicities (and later times)in the Galaxy.
Finally,
the detection of thorium and uranium in halo and globular cluster stars
offers a promising, independent age-dating technique that can put lower limits
on the age of the Galaxy and thus the Universe.}

\section{Introduction}

\begin{figure}[ht]
%\epsfxsize=10cm   %width of figure - will enlarge/reduce the figures
%\epsfbox{fig3.eps}
%\figurebox{2cm}{3cm}{} %to have a box alone 
\centerline{\epsfxsize=3.9in\epsfbox{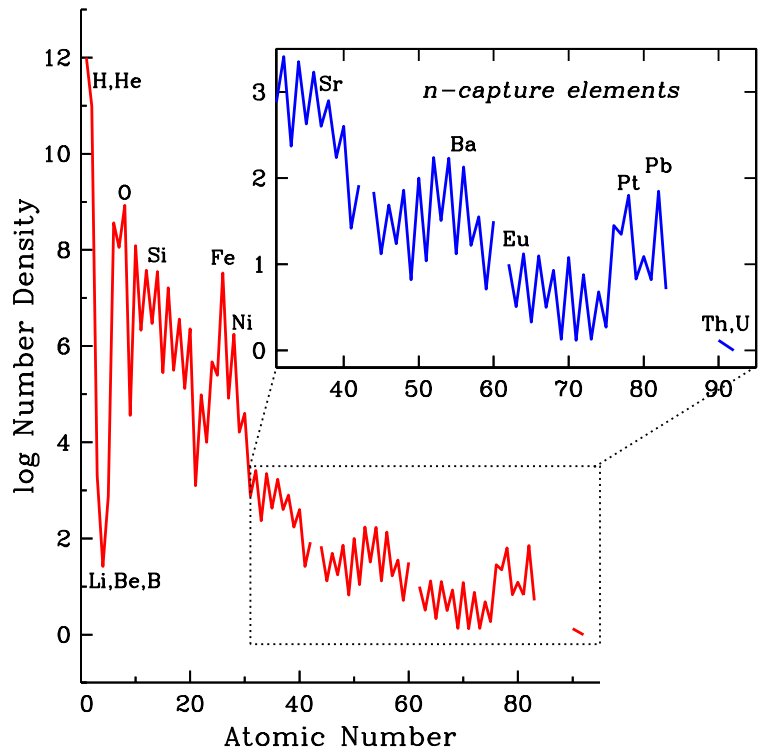}}   
\caption{
Abundances of elements in the Sun and in undifferentiated solar system material
[3]. 
This abundance set is normalized by convention to log N(H) = 12 in
astronomical literature on stars (planetary and meteoritic studies usually
normalize the abundances to log~N(Si)~= 6).
The main figure shows the entire set of stable and long-lived radioactive
elements, while the inset is restricted to the $n$-capture elements,
defined here as those elements with Z~$>$ 30.
\label{solar}}
\end{figure}

The heavy solar system abundances (here, Z $>$ 30) are formed in 
neutron capture ({\it n}-capture) processes, either the slow ({\it s}-) or 
rapid ({\it r}-) process.   
We show in Figure~\ref{solar} the solar system -- also thought of as 
cosmic -- abundances with the 
neutron-capture elements highlighted. Abundance observations of these
elements in halo stars contain vital clues  to the 
nucleosynthesis history and chemical evolution of the Galaxy,  
and abundances  of radioactive elements can also be utilized to obtain 
age determinations  for the oldest stars, which in turn put lower limits 
on age estimates for the Galaxy and the Universe ({\it e.g.}, [1,2]). 

\section{Neutron-Capture Abundances in Metal-Poor Halo Stars}

\begin{figure}[ht]
%\epsfxsize=10cm   %width of figure - will enlarge/reduce the figures
%\epsfbox{fig3.eps}
%\figurebox{2cm}{3cm}{} %to have a box alone
\centerline{\epsfxsize=3.9in\epsfbox{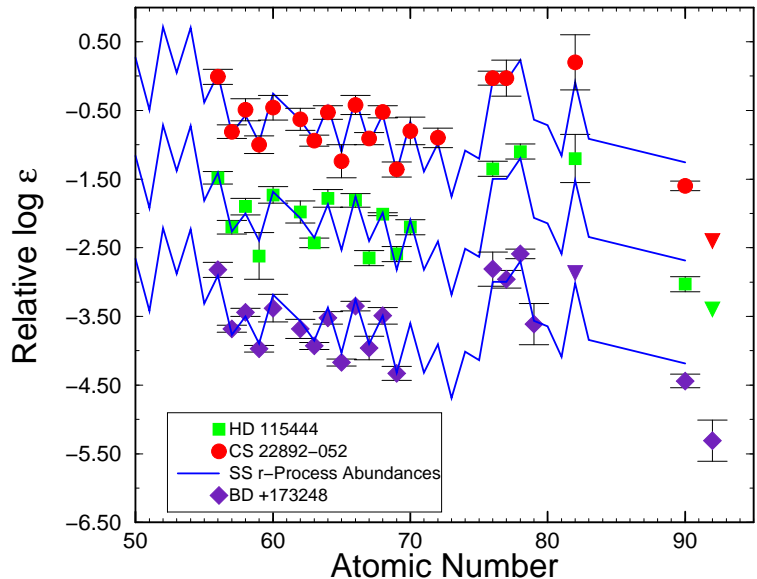}}
\caption{
A comparison of observed abundances 
in  the three stars CS~22892-052, HD~115444 and \bd17  with
a solar system  {\it r}-process 
elemental abundances. Upper limits are indicated 
by inverted triangles.
\label{three}}
\end{figure}

Extensive abundance studies have been made for metal-poor ({\it i.e.}, low iron
abundance) Galactic halo studies [4--7]. We show in  
Figure~\ref{three} 
detailed abundances
for the heaviest Z $>$  30 {\it n}-capture abundances in three stars:
CS~22892--052, \bd17 and HD~115444 [6--8]. The abundances in these stars  have been
compared to a scaled solar system curve, indicated by the solid line.

This  solid-line curve was obtained  
by summing the individual isotopic contributions from the {\it s}- and the
{\it r}-process in solar system material 
from $n$-capture cross section
measurements [9,10].
These solar  elemental {\it r}-process abundances  
that were derived are based upon the 
``classical approach'' to the $s$-process, which  is
empirical and by definition  model-independent. Nevertheless,
more sophisticated stellar models based upon
$s$-process nucleosynthesis in low-mass AGB stars,
predict very similar solar abundances [11].
As Figure~\ref{three}  illustrates there is a remarkable agreement between the 
solar {\it r}-process abundances and the abundance patterns of the heaviest 
{\it n}-capture elements in these very old metal-poor halo stars. This agreement
implies a robust {\it r}-process, one that seems to reproduce the same relative
{\it n}-capture abundances over many Gyr in the history of the Galaxy. These 
results  
further suggest a very narrow range of astrophysical conditions and/or 
mass ranges for the site(s) for the 
{\it r}-process. It has been suggested, for instance,  that only low mass 
(8--10 \Msun) supernovae
may be a likely site  for the main {\it r}-process 
and may be 
responsible for the synthesis of these heavy (Z $\ge$ 56)
{\it n}-capture elements [12--14].

This agreement between elemental abundances and the solar system {\it r}-process
abundance
distribution  has now been extended to the isotopic level. 
Recent results demonstrate that Eu isotopic abundance fractions 
in the very metal-poor,
{\it n}-capture-rich giant stars CS~22892--052, HD~115444 and \bd17 are
in excellent agreement with each other and with their values in the solar
system [15]. The Ba isotopic abundance fractions 
in one metal-poor star are also consistent with the solar {\it r}-process values
[16], again suggesting the universality of the (main) 
{\it r}-process for the
heaviest {\it n}-capture elements.  

\begin{figure}[ht]
%\epsfxsize=10cm   %width of figure - will enlarge/reduce the figures
%\epsfbox{fig3.eps}
%\figurebox{2cm}{3cm}{} %to have a box alone
\centerline{\epsfxsize=3.9in\epsfbox{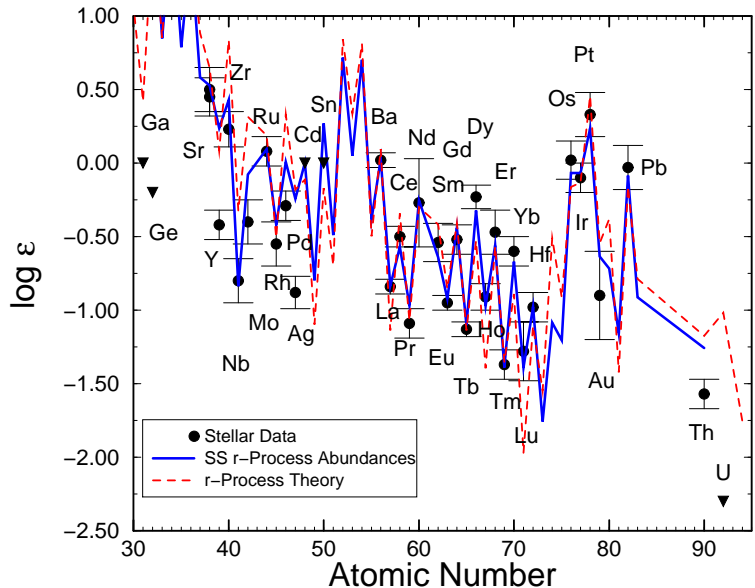}}   
\caption{
A comparison of new observed abundances from CS~22892-052 with
solar system [10] and theoretical [5] {\it r}-process elemental abundances.
\label{22sol4b}}
\end{figure}

Until recently there have been little data available on the lighter 
{\it n}-capture 
elements, particularly those 
from Z = 40--50. Recent abundances studies of \bd17 indicated that some of the 
elements in this regime, specifically  the element Ag, seemed to deviate from
the same solar curve that fit the heavier {\it n}-capture elements. 
This seemed to support an earlier suggestion, based upon solar system
meteoritic studies, of two {\it r}-processes - one for the  elements 
A $\gtaprx$ 130-140
and a second {\it r}-process for the lighter  elements
[17]. 
We show in Figure~\ref{22sol4b} preliminary new data on CS~22892--052. 
we note the detection of elements never seen before in this star, including 
Mo, Lu, Au, Pt and Pb [18].
Also significant new upper limits have been found for Ga, Ge, Cd and
Sn.  
Comparison of the abundances with the solar {\it r}-process curve
[10] demonstrates  the same agreement found previously for the heaviest
{\it n}-capture elements in this star and other similar stars
[6,1,2]. It is clear from Figure~\ref{22sol4b} that some of 
the lighter elements between Z = 40--50 ({\it e.g.}, Ag) 
show significant
deviations while others appear to fall near the line.
This new result is  consistent with earlier studies of this star and
with that of \bd17. However, with only a few stars and very limited 
data available, it is not clear at this point 
what is the source of the synthesis
for  these lighter elements. 
It has been suggested that perhaps, analogously to the {\it s}-process,
the lighter elements might be synthesized in a ``weak'' {\it r}-process 
with the heavier elements synthesized in the more robust ``strong'' or
``main'' {\it r}-process [19]. While a second {\it r}-process site,  perhaps 
supernovae of a a different mass range or frequency [20] or perhaps 
the helium zone of an exploding supernovae [19],   
might be responsible
for the synthesis of nuclei with A
$\ltaprx$ 130--140,
there have also been suggestions 
that the entire abundance distribution
could be synthesized in a single  core-collapse
supernova [6,21].

\section{Abundance Scatter and the Chemical Evolution of the Galaxy}

\begin{figure}[ht]
%\epsfxsize=10cm   %width of figure - will enlarge/reduce the figures
%\epsfbox{fig3.eps}
%\figurebox{2cm}{3cm}{} %to have a box alone
\centerline{\epsfxsize=3.6in\epsfbox{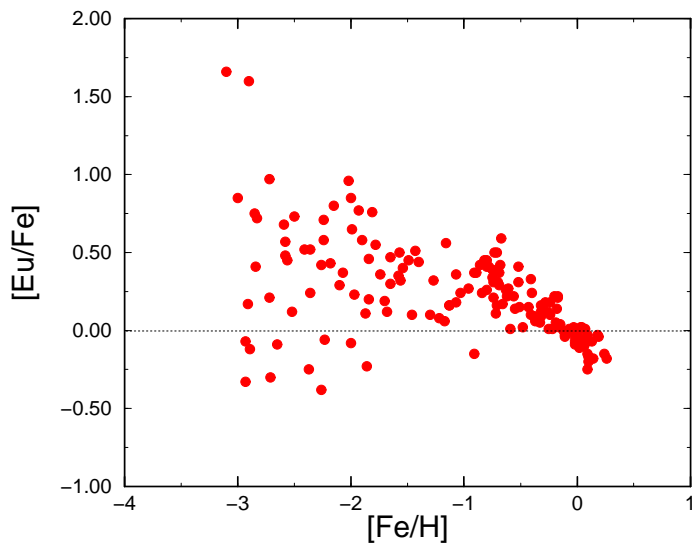}}
\caption{
The ratio [Eu/Fe] is plotted  
as a function of [Fe/H]
from various surveys of halo and disk stars [1,2].
The dotted line represents the
solar value.\label{eu7n}}
\end{figure}

A number of  studies have demonstrated a dramatic and very large 
star-to-star scatter in the abundance level of the heavy {\it n}-capture 
elements
with respect to the to iron abundances. 
This star-to-star scatter increases dramatically with decreasing stellar
metallicity, as shown in Figure~\ref{eu7n}, in which we plot
[Eu/Fe]\footnote{[A/B]~$\equiv$ log[(N$_A$/N$_B$)$_{star}$/(N$_A$/N$_B$)$_{Sun}$]} 
abundance ratios as a function of [Fe/H]
metallicity for a large
number of halo and disk stars [1,2]. Eu is employed for such studies since
it is relatively easy to detect in the spectra of metal-poor stars and 
because it is predominantly produced in the {\it r}-process.
It is seen in Figure~\ref{eu7n} that near [Fe/H] = --3,   
(which includes some of the oldest
stars in the Galaxy), the [Eu/Fe] ratio reaches a  peak of
$\sim$ 50 and varies from star-to-star by more than a factor of 100.
Thus, even though the [Eu/H] ratios are still less than that of the sun,
the relative ratio of this {\it r}-process element to iron in some of these
stars is much larger than the solar ratio.
It is also clear from the figure that this star-to-star scatter decreases 
with increasing [Fe/H] 
tending toward younger (on average) stars.
While there could be several possible explanations for this scatter, the most
likely  interpretation is   that the early Galaxy was
chemically inhomogeneous and unmixed.

Abundance trends with metallicity also demonstrate the ongoing chemical 
evolution of the Galaxy. In the past [Ba/Eu] has been employed to study
the changing contributions to the {\it r}- and {\it s}-processes. 
The element Ba is overwhelmingly synthesize in the {\it s}-process and 
provides a good indication of the $s$-process $n$-capture
nucleosynthesis history of the Galaxy.
There is,  however, significant observationally related 
scatter in the Ba abundance data making it 
less reliable for such studies.  As an alternative,  new studies are now
employing La,   
also a  predominantly {\it s}-process element,  to Eu to examine 
Galactic chemical evolution.
Very preliminary results [22] seem to indicate some  initial indication  of the 
{\it s}-process below [Fe/H] = --2 
with the major onset  of the ejection of this material into the Galaxy 
occurring near 
a metallicity of --2. These  results lend support to earlier studies 
with similar findings [10,23].

\section{Radioactive Chronometers and the Age of the Galaxy}

Abundance detections of certain long-lived radioactive isotopes can be 
employed as ``chronometers'' or clocks to determine the ages of the oldest
stars. There have been a number of recent detections of the element thorium,
with a half-life of 14 Gyr, in the metal-poor halo halo stars [4--8,24,25]. 
This element, along with uranium, is synthesized solely in the {\it r}-process.
Comparison of the observed stellar abundance of this radioactive element
with its initial (time-zero) abundance in an {\it r}-process site leads to a direct
radioactive-age estimate of the star.  
We show in  Figure~\ref{22sol4b} the abundance distribution, including Th,
in the ultra-metal-poor star
CS~22892--052. While (as noted before) the heavy {\it n}-capture elements are 
consistent with the scaled solar {\it r}-process curve, the observed Th 
abundance lies
below this same line. This difference is a clear demonstration that this star 
is older than the sun. To determine how much older requires knowledge of 
the initial Th abundance that must be predicted from {\it r}-process models. 
Such a model calculation is illustrated in Figure~\ref{22sol4b} 
by the dashed line [5]. 
The goal of 
such theoretical calculations is to reproduce the stellar, and hence the 
solar system {\it r}-process, abundance pattern and at the same time predict the 
abundances of the radioactive elements. Such predictions, to reduce errors,
employ the ratio of Th to another {\it r}-process element, typically Eu. 
Utilizing this technique has led to chronometric age estimates ranging from
$\simeq$ 11--15 $\pm$ 4 Gyr [4--8,24,25] that are consistent with 
globular cluster ages and cosmological age estimates based upon the 
observed supernova expansion rates. 
However,  the Th/Eu chronometer gives a very different and completely inconsistent
age in the star CS~31082--001 [25]. This star was the first with a detection of 
U and the Th/U chronometer does give  an age of 15.5 Gyr [26]. Since Th/Eu and
Th/U give similar results in \bd17 [7], it is not
clear yet why CS~31082--001 is so different or if it is rare.
Clearly additional U detections will be needed to answer this question.
Just as importantly reliable nuclear data, experimental where available but 
mostly theoretical predictions, for the neutron-rich nuclei in the 
{\it r}-process 
will be necessary to determine more accurately the ages of these old stars
and put limits on the age of the Galaxy and the Universe.

\section*{Acknowledgments}

Partial support for this research was provided by the National
Science Foundation
(AST-9986974  to JJC and AST-9987162  to CS)
and by the Space Telescope Science grant GO-08342. 
JJC thanks the University of Texas at Austin Department of
Astronomy John W. Cox Fund for partial support while this
paper was being written.

%\par\vfil\eject
\end{document}